\documentclass[ %
  10pt,
  twocolumn,
  aps,
  prb,                      
  groupedaddress,
  superscriptaddress,
  amsmath,amssymb,
  showkeys,                 
  showpacs,                 
  noeprint,                 
  english,
  longbibliography 
]{revtex4-2}

\usepackage[T1]{fontenc}  
\usepackage{times}
\usepackage[english]{babel}                
\usepackage{multirow}
\usepackage{tikz}
\usetikzlibrary{arrows.meta,positioning,fit,shapes.geometric}
\usepackage{array}
\usepackage[section]{placeins}
\usepackage{graphicx}                      
\usepackage{dcolumn}                       
\usepackage{bm}                            
\usepackage{mathrsfs,amsbsy,amstext}       
\usepackage{setspace}                      
\usepackage{esint}                         
\usepackage{booktabs}                      
\usepackage{tikz}                          
\usepackage{braket}                        
\usepackage{enumitem}
\usepackage{listings}
\usepackage{tabularx}
\usepackage{booktabs}
\usepackage{siunitx}
\usepackage{xurl}
\usepackage[normalem]{ulem}

\usetikzlibrary{%
  positioning,   
  arrows.meta    
}

\usepackage[colorlinks,
            linkcolor=blue,
            citecolor=blue,
            urlcolor=blue,
            hypertexnames=true]{hyperref}

\begin{document}

\title{Measurement-Guided State Refinement for Shallow Feedback-Based Quantum Optimization Algorithm}


\author{Lucas A. M. Rattighieri}
\thanks{These authors contributed equally to this work.}
\affiliation{Instituto de Física Gleb Wataghin, Universidade de Campinas, Campinas, 13083-970, SP, Brazil}

\author{Pedro M. Prado}
\thanks{These authors contributed equally to this work.}
\affiliation{S\~ao Paulo State University (UNESP), School of Sciences,
            17033-360 Bauru-SP, Brazil}

\author{Marcos C. de Oliveira}
\affiliation{Instituto de Física Gleb Wataghin, Universidade de Campinas, Campinas, 13083-970, SP, Brazil}
\affiliation{QuaTI – Quantum Technology \& Information,
            13560-161 S\~ao Carlos-SP, Brazil}

\author{Felipe F. Fanchini}
\email{felipe.fanchini@unesp.br}
\affiliation{Hospital Israelita Albert Einstein, São Paulo-SP, Brazil}
\affiliation{S\~ao Paulo State University (UNESP), School of Sciences,
            17033-360 Bauru-SP, Brazil}
            
\date{\today}



\begin{abstract}
Limited circuit depth remains a central constraint for quantum optimization
in the noisy intermediate-scale quantum (NISQ) regime, where shallow
unitary dynamics may fail to sufficiently concentrate probability on
low-energy configurations. We introduce Measurement-Guided Initialization
(MGI), an iterative strategy that uses measurement outcomes from previous
executions to update the initialization of subsequent runs. The method
extracts single-qubit marginal probabilities from dominant measurement
outcomes and prepares a biased product-state initialization, allowing
information obtained during optimization to be reused without introducing
classical parameter optimization. We implement this approach in the
context of the Feedback-Based Algorithm for Quantum Optimization
(FALQON) and evaluate its performance on weighted \textit{MaxCut}
instances. Numerical results show that measurement-guided initialization
improves the performance of shallow-depth circuits and enables iterative
refinement toward high-quality solutions while preserving the
non-variational structure of the algorithm. These results indicate that
measurement statistics can be exploited to improve shallow quantum
optimization protocols compatible with NISQ devices.
\end{abstract}

\maketitle

\section{Introduction}

Combinatorial optimization problems play a central role in science and technology, with applications ranging from logistics and networks to bioinformatics \cite{Ahuja1993,KorteVygen2018,Schrijver2003,Gusfield1997}. The intrinsic difficulty of many such tasks, often NP-hard, motivates the search for quantum algorithms capable of exploiting problem structure and quantum resources to produce high-quality solutions \cite{Lucas2014,Glover2018}. However, in the current era of noisy intermediate-scale quantum (NISQ) hardware, realizing this potential remains a significant challenge. Severe limitations on coherence times and noise sensitivity strictly constrain the feasible circuit depth, compelling the development of shallow and noise-resilient optimization protocols \cite{Preskill2018,Cerezo2021}.

Despite these hardware limitations, most quantum optimization algorithms rely on ansatz architectures that can become prohibitively deep and, consequently, highly susceptible to noise. These methods generally employ an alternating sequence of operators, applying a problem Hamiltonian followed by a mixer, a structure common to the Quantum Approximate Optimization Algorithm (QAOA) \cite{Farhi2014,Hadfield2019}, Trotterized adiabatic evolution, and the \textit{Feedback-Based Algorithm for Quantum Optimization} (FALQON) \cite{Magann_2022,FQA}. The challenges associated with these approaches, however, manifest differently. In QAOA, circuit depths can be kept relatively shallow, but the parameters must be determined through a classical optimization loop that is often costly and prone to inefficiencies. In contrast, FALQON bypasses classical optimization by updating parameters layer-by-layer based on measurement feedback, guaranteeing a monotonic decrease in the cost function energy. Nevertheless, achieving convergence typically requires deep circuits, which renders its implementation impractical on present-day noisy devices.

In this context, several approaches and variations have been proposed to accelerate the convergence of FALQON and reduce the required circuit depth. Among these, SO-FALQON \cite{Arai2025} employs an alternative feedback law based on a second-order approximation of the evolution operators, enabling more informative parameter updates. Strategies inspired by counterdiabatic control \cite{PhysRevResearch.6.043068,chandarana2024lyapunovcontrolledcounterdiabaticquantum} have also been introduced, aiming to suppress unwanted transitions and accelerate the preparation of the target state. Other approaches explore temporal rescaling techniques that modify the evolution schedule to improve convergence efficiency \cite{Rattighieri2025}. Additionally, methods such as FOCQS \cite{FOCQS} refine the control parameters of previous layers using perturbative techniques, incorporating corrections that enhance the overall performance of the algorithm. Although these strategies demonstrate improved convergence behavior, the number of layers required to reach high-quality solutions generally remains large, limiting their practical applicability on NISQ devices. 

A key limitation shared by these approaches is that, while feedback determines the dynamical evolution of the circuit, the initialization of the algorithm typically remains structure-agnostic, most commonly starting from a uniform superposition state. As a consequence, information obtained from previous executions is not reused to guide subsequent initializations. This observation motivates the following question: is it possible to retain the feedback-driven advantages of FALQON while strictly limiting the circuit to a shallow, non-optimized structure? 

To address this question, we introduce \textit{Measurement-Guided Initialization} (MGI) for FALQON (MGI-FALQON). The proposed strategy incorporates an iterative outer loop that refines the starting point of the algorithm using data obtained from previous shallow executions. Specifically, MGI extracts single-qubit marginal probabilities from a filtered subset of the most frequent output bitstrings. These marginals are then used to define a biased product state, prepared via local $R_y$ rotations, which guides the algorithm toward high-quality solutions without increasing circuit depth or introducing classical parameter optimization. In this way, information contained in measurement statistics is iteratively reused to concentrate probability mass on promising regions of the search space while preserving the feedback-driven and non-variational character of FALQON.

As a case study, we employ \textit{MaxCut}, a canonical benchmark in combinatorial optimization widely used to evaluate quantum algorithms \cite{Lucas2014,Hadfield2019}. This choice is motivated by its universality, since any Quadratic Unconstrained Binary Optimization (QUBO) problem can be mapped to a weighted MaxCut instance \cite{Barahona1989, docarmo2025}, making MaxCut a natural proxy for assessing progress toward solving general combinatorial optimization tasks on NISQ hardware.

During the development of this work, a closely related approach was independently introduced in the context of a non-variational version of QAOA, referred to as Iterative-QAOA \cite{lopezruiz2025nonvariationalquantumapproachjob}. In that method, the circuit structure and parameter schedule are kept fixed, while the initial state is iteratively refined based on measurement outcomes from previous runs. Similar to our approach, the update procedure constructs a biased product state using single-qubit marginal information extracted from measured bitstrings and prepares it through local $R_y$ rotations. However, the update rule differs in that Iterative-QAOA computes qubit biases using a Boltzmann-weighted average over the measured energies, introducing a temperature-dependent feedback mechanism that emphasizes low-energy configurations. This conceptual similarity reinforces the relevance of measurement-guided initialization strategies as a broader framework for enhancing shallow quantum optimization algorithms, and highlights the growing interest in iterative, measurement-driven protocols that improve performance without increasing circuit depth or relying on classical parameter optimization. In this way, MGI should be viewed as a measurement-driven initialization framework that can be combined with different feedback-based or variational optimization schemes.

The paper is organized as follows. In Section~\ref{sec:maxcut_qubo}, we introduce the \textit{MaxCut} problem and its QUBO formulation, defining the cost function and notation. Section~\ref{sec:background_falqon} reviews FALQON. Section~\ref{sec:mgi} details the proposed method, \textit{Measurement-Guided Initialization} for FALQON, and the measurement-based feedback loop. Section~\ref{sec:setup} outlines the experimental protocol, including instances, metrics, and resource budgets. Finally, results are discussed in Section~\ref{sec:results}, followed by our conclusions in Section~\ref{sec:conclusions}.

\section{The MaxCut Problem and QUBO Formulation}\label{sec:maxcut_qubo}

\begin{figure}[t]
    \centering
    \includegraphics[width=0.9\linewidth]{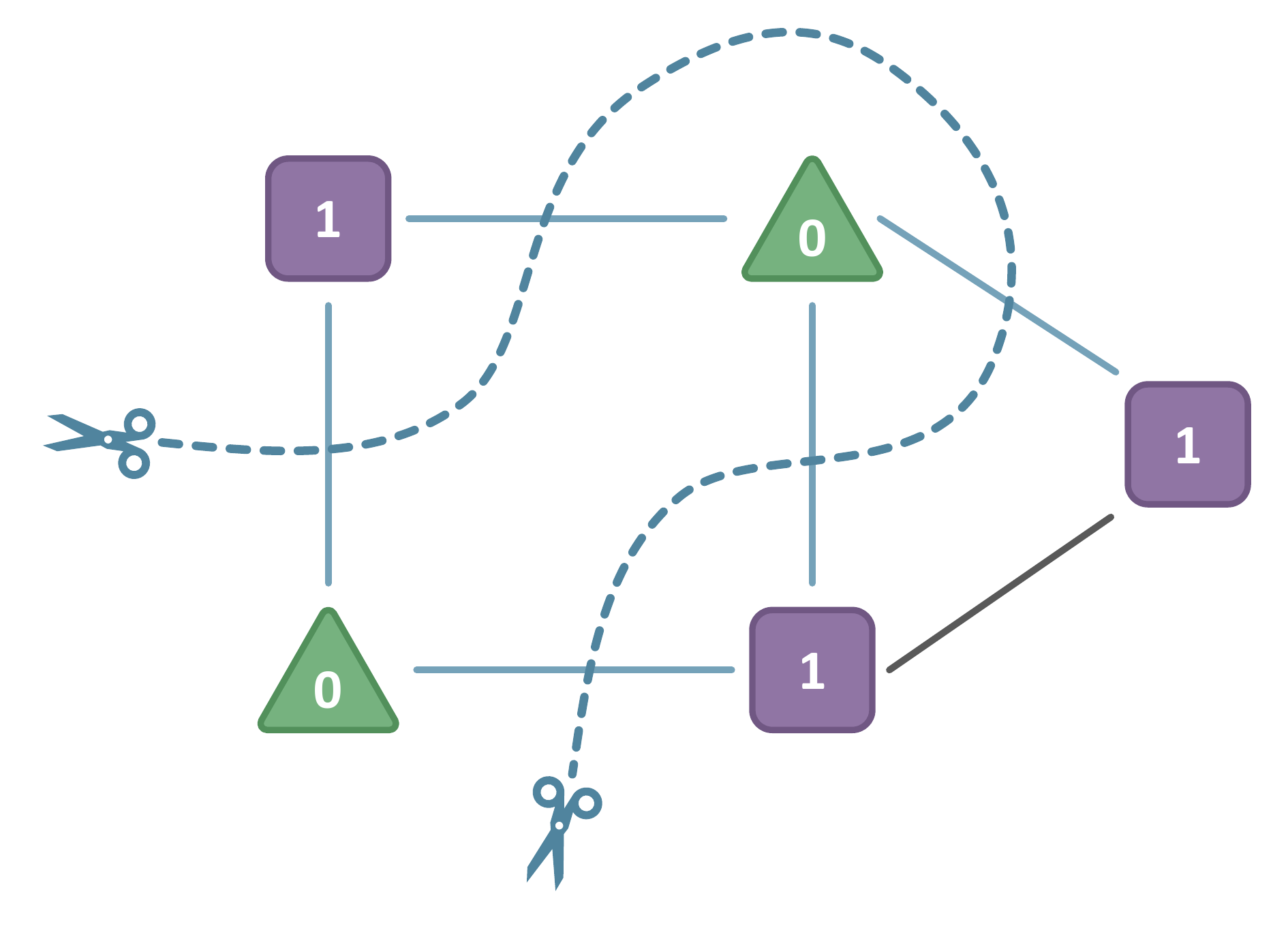}
    \caption{Schematic representation of the MaxCut problem on a graph. Vertices are partitioned into two subsets, where purple squares represent vertices encoded in the state $|1\rangle$ and green triangles represent vertices encoded in the state $|0\rangle$. The objective is to maximize the total weight of the edges crossing the partition (highlighted in blue), corresponding to the contribution captured in Eq.~\eqref{eq:cut_set}.}
    \label{fig:maxcut}
\end{figure}

Given a weighted graph $G=(V,E,w)$ with non-negative edge weights $w_{ij}\ge 0$, the \textit{MaxCut} problem seeks a bipartition of the vertex set $V$ into two disjoint subsets, $S$ and $\bar{S}$, such that the total weight of edges connecting $S$ to $\bar{S}$ is maximized. Formally, the cut value is defined as
\begin{equation}
\mathrm{cut}(S) = \sum_{(i,j)\in E} w_{ij} \cdot \mathbb{I}\left[ (i \in S \land j \in \bar{S}) \lor (i \in \bar{S} \land j \in S) \right],
\label{eq:cut_set}
\end{equation}
where $\mathbb{I}[\cdot]$ denotes the indicator function. Figure~\ref{fig:maxcut} illustrates this concept, highlighting the edges that contribute to the objective function.

To obtain an algebraic formulation suitable for optimization, we introduce binary variables $x_i \in \{0,1\}$ for each vertex $i$, where $x_i=1$ if $i \in S$ and $x_i=0$ otherwise. The condition that an edge $(i,j)$ crosses the cut corresponds to the exclusive-OR (XOR) operation, which can be written as $x_i \oplus x_j = x_i + x_j - 2x_i x_j$. The classical objective function can therefore be expressed as
\begin{equation}
C(x) = \sum_{(i,j)\in E} w_{ij}\,(x_i + x_j - 2x_i x_j).
\label{eq:cut_qubo_scalar}
\end{equation}

Equation~\eqref{eq:cut_qubo_scalar} admits a matrix representation in the form of a Quadratic Unconstrained Binary Optimization (QUBO) problem,
\begin{equation}
C(x) = x^\top Q\,x,
\label{eq:qubo_matrix_form}
\end{equation}
where the symmetric matrix $Q$ encodes the structure of the graph. A standard construction yields
\begin{align}
Q_{ij} &=
\begin{cases}
\sum_{k \in \mathcal{N}(i)} w_{ik} & \text{if } i=j, \\[4pt]
-2\,w_{ij} & \text{if } (i,j)\in E, \\[4pt]
0 & \text{otherwise,}
\end{cases}
\label{eq:Q}
\end{align}
where $\mathcal{N}(i)$ denotes the neighborhood of vertex $i$. Maximizing $C(x)$ is therefore equivalent to solving $\arg\max_{x\in\{0,1\}^n} x^\top Q\,x$.

For quantum optimization, it is convenient to map binary variables to spin variables $z_i \in \{-1, +1\}$ through the transformation $z_i = 1 - 2x_i$. Substituting this relation into Eq.~\eqref{eq:cut_qubo_scalar}, the objective function takes the form of an Ising model,
\begin{equation}
C(z) = \frac{1}{2}\sum_{(i,j)\in E} w_{ij}\,(1 - z_i z_j),
\label{eq:ising_cost}
\end{equation}
which naturally leads to a quantum Hamiltonian formulation. Promoting the spin variables to quantum operators ($z_i \to Z_i$), we obtain the cost Hamiltonian diagonal in the computational basis,
\begin{equation}
H_p = \frac{1}{2}\sum_{(i,j)\in E} w_{ij}\,(\mathbb{I} - Z_i Z_j),
\label{eq:hamiltonian_cost}
\end{equation}
where $Z_i$ denotes the Pauli-$Z$ operator acting on qubit $i$. The expectation value $\langle H_p \rangle$ directly corresponds to the average cut value. Since the operator consists exclusively of diagonal $Z$ terms, its energy can be efficiently estimated by sampling bitstrings in the computational basis, a feature that plays a central role in the \textit{Measurement-Guided Initialization} strategy proposed in this work.

The identity terms contribute only a uniform energy shift and therefore do not affect the eigenstates of the Hamiltonian. Similarly, the global factor $1/2$ and the edge weights $w_{ij}$ rescale eigenvalues without modifying eigenvectors. Because the optimization procedure depends only on the relative ordering of energies, these constant and multiplicative factors can be omitted without loss of generality. The cost Hamiltonian can thus be written in the simplified form
\begin{equation}
H_p = - \sum_{(i,j)\in E} w_{ij}\, Z_i Z_j,
\label{eq:hamiltoniano_simplificado}
\end{equation}
which preserves the same eigenstates and encodes the MaxCut structure entirely through pairwise correlations between qubits.

A key conceptual aspect of this formulation is the global $\mathbb{Z}_2$ bit-flip symmetry. Swapping the subsets $S$ and $\bar{S}$ (i.e., applying the transformation $x_i \to 1-x_i$ or $z_i \to -z_i$ for all $i$) leaves the cut value invariant. In the quantum formulation, this implies that the Hamiltonian commutes with the global bit-flip operator $\Pi_i X_i$, resulting in a degeneracy where every solution possesses a symmetry-related counterpart.

To remove this redundancy, a reference vertex is fixed, chosen as vertex $0$, and assigned to the subset labeled by $0$. In the quantum formulation, this corresponds to fixing the associated qubit in the state $\ket{0}$. The total quantum state can then be written as $\ket{\Psi} = \ket{0}_0 \otimes \ket{\psi}_{\mathrm{rest}}$. Since $\ket{0}_0$ is an eigenstate of $Z_0$, the Hamiltonian does not induce transitions on this qubit, and the dynamics is fully determined by an effective Hamiltonian acting on the remaining qubits. The fixed qubit can therefore be removed from the simulation without loss of generality.

In practice, this is implemented by starting from Eq.~\eqref{eq:hamiltoniano_simplificado} and fixing $Z_0 = +1$, yielding
\begin{equation}
H_p= -\sum_{\substack{(i, j) \in E \\ i, j \neq 0}}  w_{i j} Z_i Z_j
-\sum_{j:(0, j) \in E} w_{0 j}Z_j.
\label{eq:hamiltoniano_vertice_fixado}
\end{equation}
This reduced Hamiltonian acts on a Hilbert space with one fewer qubit while preserving the same optimal solutions as the original problem, with the symmetry-related redundancy removed.

Finally, the choice of MaxCut is strategically motivated by its universality. Since any general QUBO problem can be mapped to a weighted MaxCut instance \cite{Barahona1989}, it provides a robust proxy for the broader class of combinatorial optimization tasks. Moreover, the corresponding cost Hamiltonian requires only diagonal $Z$-basis measurements, simplifying the collection of bitstring statistics and the accounting of shot budgets. These features make MaxCut particularly suitable for evaluating feedback-based optimization algorithms such as FALQON and for assessing the impact of \textit{Measurement-Guided Initialization} in shallow-depth regimes.

\section{FALQON}\label{sec:background_falqon}

The \textit{Feedback-Based Algorithm for Quantum Optimization} (FALQON) is a constructive protocol inspired by quantum Lyapunov control theory~\cite{Magann_2022}. Unlike variational algorithms that rely on classical optimization loops to explore a parameter landscape, FALQON builds the circuit sequentially, layer-by-layer. At each step, the parameter associated with the new layer is determined through a feedback rule derived from measurements of the current quantum state. This strategy guarantees a monotonic decrease of the cost function energy, systematically guiding the system toward lower-energy configurations as the circuit depth increases.

Consider a problem described by a Hamiltonian $H_p$, whose ground state encodes the solution of interest, and a driver (or mixer) Hamiltonian $H_d$, typically chosen such that its ground state is easy to prepare. The algorithm constructs a depth-$L$ circuit as a sequence of Trotterized time-evolution operators. Let $|\psi_{\ell-1}\rangle$ denote the state at layer $\ell-1$. The state at layer $\ell$ is obtained as
\begin{equation}
|\psi_{\ell}\rangle = U_d(\beta_\ell \Delta t)\, U_p(\Delta t)\, |\psi_{\ell-1}\rangle,
\label{eq:falqon_update}
\end{equation}
where $U_p(\tau)=e^{-i H_p \tau}$ and $U_d(\tau)=e^{-i H_d \tau}$. The time step $\Delta t > 0$ is a fixed hyperparameter, while $\beta_\ell$ is a variable parameter determined by the feedback law. The procedure begins from an initial state $|\psi_0\rangle$, commonly chosen as the ground state of $H_d$.

The central mechanism of FALQON is the assignment of $\beta_\ell$. To ensure that the energy $E_\ell = \langle \psi_\ell | H_p | \psi_\ell \rangle$ decreases (or remains constant) at each step, the parameter is chosen proportional to the time derivative of the energy expectation value. This leads to a feedback rule expressed in terms of the commutator between the Hamiltonians,
\begin{equation}
\beta_{\ell} = -\alpha \langle i[H_d,H_p]\rangle_{\ell-1},
\label{eq:feedback_rule}
\end{equation}
where $\langle \cdot \rangle_{\ell-1}$ denotes the expectation value with respect to the state $|\psi_{\ell-1}\rangle$, and $\alpha$ is a gain factor that can be absorbed into $\Delta t$ or set to unity in simplified implementations.

In experimental realizations, the value of $\beta_\ell$ is obtained by measuring the observable $\mathcal{O} = i[H_d, H_p]$ on the current state. As a consequence, the circuit depth increases deterministically: measurements performed at layer $\ell-1$ directly determine the control parameter applied at layer $\ell$. This construction eliminates the classical optimization overhead characteristic of variational algorithms and mitigates issues such as barren plateaus associated with random parameter initialization. However, because the evolution follows a strictly feedback-driven Lyapunov trajectory, convergence toward low-energy states typically requires a large number of layers~\cite{Magann_2022,Pexe2024}. This requirement persists even when considering recent improvements aimed at accelerating convergence or modifying the feedback dynamics \cite{Rattighieri2025,Arai2025}.

As a result, although FALQON provides a robust non-variational framework for quantum optimization, its practical applicability on NISQ devices remains limited by the circuit depth required to reach high-quality solutions. In the next section, we address this bottleneck by introducing a strategy that uses measurement outcomes from shallow executions to iteratively refine the initialization of the algorithm, reducing the required circuit depth while preserving the non-parameterized and feedback-driven character of FALQON.

\section{Measurement-Guided Initialization (MGI) for FALQON}\label{sec:mgi}

\begin{figure*}[t]
\centering
\includegraphics[width=\linewidth]{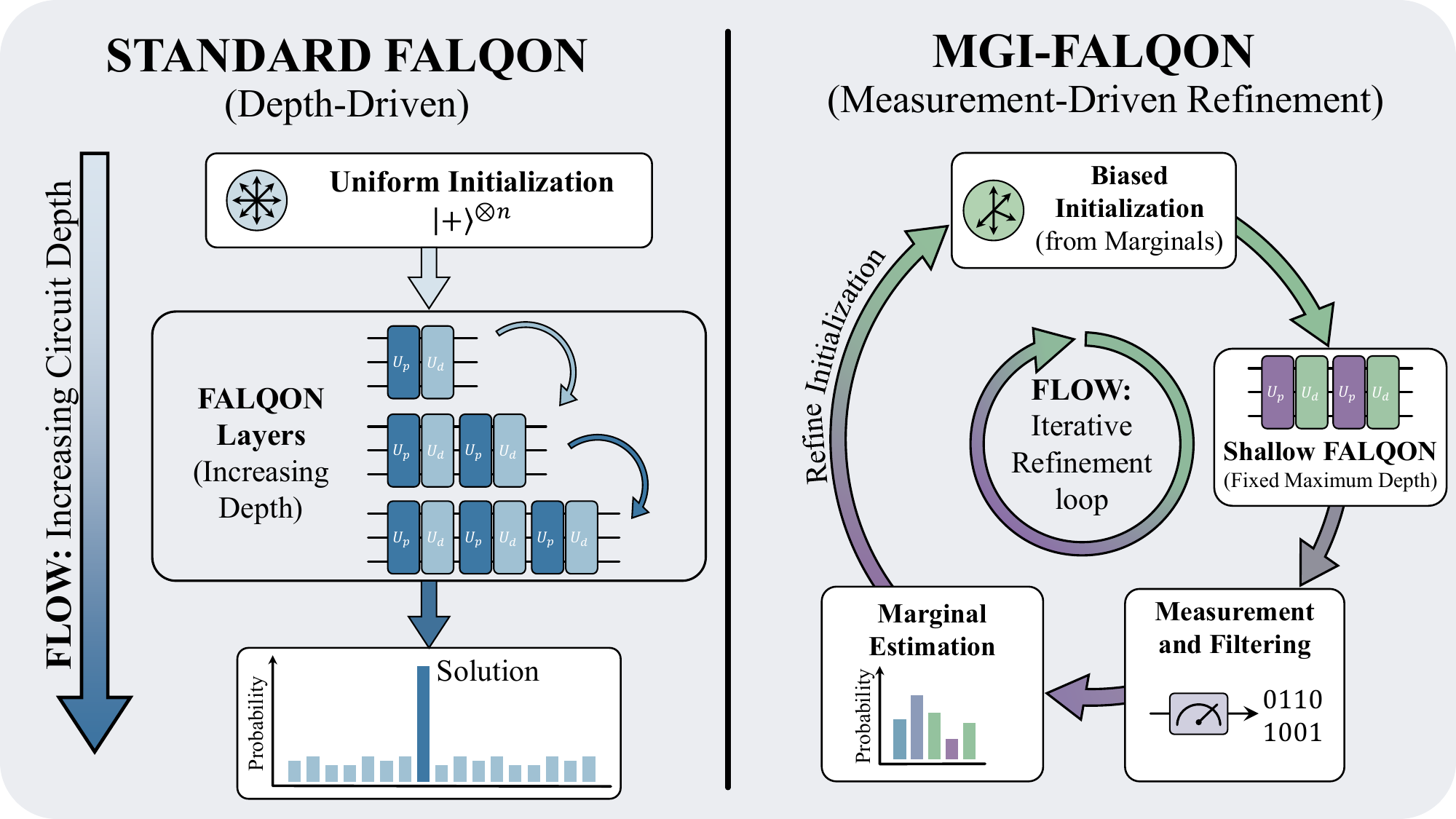}
\caption{
Conceptual comparison between standard FALQON and MGI-FALQON.
In standard FALQON (left), probability concentration toward low-energy configurations is achieved by increasing circuit depth.
In MGI-FALQON (right), shallow circuits are repeatedly executed and measurement outcomes are used to refine the initialization through empirical marginals, effectively replacing part of the required circuit depth by measurement-driven iterative refinement.
}
\label{fig:falqon_vs_mgi}
\end{figure*}

\begin{figure*}[t]
    \centering
    \includegraphics[width=1\linewidth]{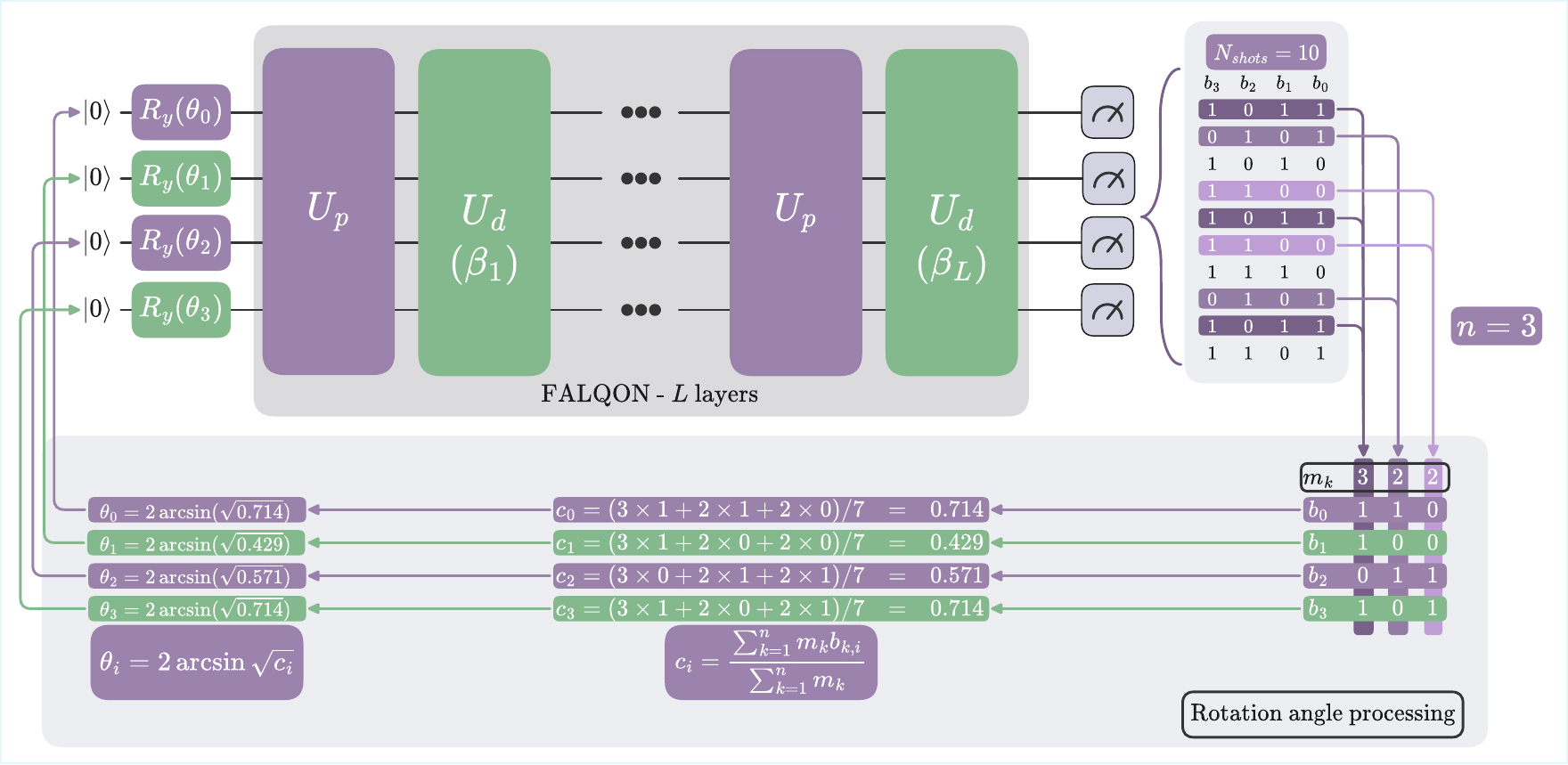}
    \caption{The diagram shows the operation of MGI-FALQON in a single iteration for a 4-qubit system. The iteration starts with the initial preparation, where each qubit receives an $R_y$ gate parametrized by the angles $\theta_i$ obtained in the previous iteration. The FALQON circuit with $L$ layers is then executed. The final state is measured $N_{\text{shots}} = 10$ times, producing 10 bitstrings. From these results, the $n = 3$ most frequent bitstrings are selected. Based on this reduced set, the probabilities $c_i$ of observing the value 1 on each qubit are computed. Finally, the angle $\theta_i$ used in the next iteration is determined so that the gate $R_y(\theta_i)$ produces a probability $c_i$ of measuring the state 1.}
    \label{fig:mgi_diagram}
\end{figure*}

In this section, we introduce the Measurement-Guided Initialization (MGI) strategy, an iterative procedure designed to improve the performance of FALQON when the circuit depth is fixed. In the standard implementation, FALQON typically starts from the uniform superposition state, which does not incorporate information about the problem structure or about outcomes obtained in previous executions. The MGI approach modifies this scheme by introducing an outer iterative loop that uses measurement information extracted from previous shallow executions to redefine the initial state of subsequent runs.
The conceptual difference between the depth-driven probability concentration of standard FALQON and the measurement-driven refinement introduced here is illustrated schematically in Fig.~\ref{fig:falqon_vs_mgi}.

The protocol is organized into $R$ iterations. At each iteration $r$, a fixed-depth FALQON circuit is executed and measured. The resulting measurement outcomes are then processed to update the parameters defining the initial state for iteration $r+1$. This update consists of two main stages: (i) a filtering step applied to the measured bitstrings, and (ii) a state-preparation procedure constructed from the estimated single-qubit marginal probabilities. In this way, information contained in measurement statistics is iteratively reused to bias the initialization toward regions of the search space that appear more frequently during the evolution. Both stages are detailed in below.
The overall workflow of a single MGI-FALQON iteration is illustrated in Fig.~\ref{fig:mgi_diagram}, which summarizes the filtering, marginal estimation, and reinitialization steps.

\subsection{Filtering and Marginal Estimation}

Consider the $r$-th iteration of the MGI loop. A FALQON circuit with fixed depth $L$ is executed and measured $N_{\text{shots}}$ times, producing a multiset of bitstrings sampled from the output distribution. In shallow-circuit regimes typical of NISQ devices, this distribution is generally broad and may retain significant weight on high-energy configurations. To reduce the influence of such configurations, we introduce a filtering step that retains only the most relevant outcomes.

The observed bitstrings are sorted by frequency, and we define the subset $\mathcal{S}_n = \{b_k\}_{k=1}^n$ containing the $n$ most frequent ones. This post-selection procedure suppresses low-frequency events, which are more likely associated with high-energy configurations or sampling noise, and concentrates the update on the dominant structures present in the distribution.

Let $m_k$ denote the multiplicity of bitstring $b_k \in \mathcal{S}_n$, and let $b_{k,i} \in \{0,1\}$ denote the value of qubit $i$ in $b_k$. From this filtered set, we estimate the empirical marginal probability that qubit $i$ is found in state $\ket{1}$,
\begin{equation}
c_i^{(r)} = \frac{\sum_{k=1}^n m_k \, b_{k,i}}{\sum_{k=1}^n m_k}, 
\qquad i=1,\dots,n_q,
\label{eq:mgi_marginals}
\end{equation}
where $n_q$ is the total number of qubits. The coefficient $c_i^{(r)}$ represents the weighted frequency with which qubit $i$ appears in state $\ket{1}$ among the most frequent outcomes. In this way, the update rule captures local biases emerging from the dominant bitstrings at iteration $r$ while ignoring the tails of the distribution. Although the energy of each bitstring is not explicitly evaluated, the procedure relies on the fact that FALQON progressively decreases the expectation value of the cost Hamiltonian, which increases the statistical weight of lower-energy configurations along the evolution.

\subsection{Biased State Preparation}

The estimated marginals directly define the initialization for the next iteration. The goal is to prepare a product state in which the probability of measuring qubit $i$ in state $|1\rangle$ matches the empirical marginal $c_i^{(r)}$. Accordingly, the quantum state for iteration $(r+1)$ is prepared as
\begin{equation}
|\psi_0^{(r+1)}\rangle = \bigotimes_{i=1}^{n_q} R_y\left(\theta_i^{(r)}\right)|0\rangle,
\label{eq:mgi_state}
\end{equation}
where the rotation angle $\theta_i^{(r)}$ is determined by
\begin{equation}
\theta_i^{(r)} = 2\arcsin\left(\sqrt{c_i^{(r)}}\right).
\label{eq:theta_update}
\end{equation}

By mapping the empirical marginals directly onto rotation angles, MGI-FALQON concentrates the probability mass of the initial state on regions of Hilbert space that appeared most frequently in previous iterations. As the iterative loop progresses, this sequence of measurement-driven refinements guides the shallow FALQON circuit toward increasingly lower-energy configurations, effectively incorporating structural information extracted from the measurement outcomes into the initialization itself while preserving the non-variational and feedback-driven nature of the algorithm.

The effectiveness of the MGI procedure can be understood from an information-theoretic perspective. After a shallow FALQON execution, the measured bitstrings define an empirical probability distribution over computational basis states that is typically broad but biased toward low-energy configurations due to the monotonic decrease of the expected energy enforced by the feedback rule. The estimation of single-qubit marginals from the most frequent outcomes can be interpreted as a projection of this empirical distribution onto the manifold of product distributions that best reproduces the observed local statistics. In this sense, the reinitialization step constructs a mean-field approximation to the dominant region of the output distribution, compressing the information contained in measurement statistics into a product-state initialization. Subsequent FALQON executions then evolve from a state with increased overlap with low-energy configurations identified in previous iterations. Iterating this process effectively replaces part of the depth required for unitary concentration of probability by a sequence of measurement-driven refinements, allowing shallow circuits to progressively concentrate probability on high-quality solutions while preserving the feedback-based and non-variational structure of FALQON.

This conceptual framework can be rigorously formalized as an iterative projection onto a product-state manifold.
Let $\hat p_r(x)$ denote the empirical output distribution at MGI iteration $r$ (after the post-selection step on $\mathcal{S}_n$). MGI replaces $\hat p_r$ by the product distribution
\begin{equation}
q_r(x)=\prod_{i=1}^{n_q} q_{r,i}(x_i),
\end{equation}
that matches the observed one-qubit statistics. Equivalently, $q_r$ is the unique factorized distribution minimizing the Kullback--Leibler divergence to $\hat p_r$,
\begin{equation}
q_r=\arg\min_{q\in\mathcal{Q}}\; D_{\mathrm{KL}}\!\left(\hat p_r\,\|\,q\right),
\qquad 
\mathcal{Q}=\left\{q(x)=\prod_{i} q_i(x_i)\right\},
\label{eq:kl_projection}
\end{equation}
whose solution satisfies $q_{r,i}(1)=c_i^{(r)}$ and $q_{r,i}(0)=1-c_i^{(r)}$. Preparing $\ket{\psi_0^{(r+1)}}$ via local $R_y$ rotations, therefore, implements a measurement-induced mean-field refinement step: information contained in the measurement statistics is compressed into a product-state initialization that increases the overlap with the dominant low-energy region sampled by shallow FALQON. Iterating this projection-and-evolution cycle effectively trades part of the circuit depth required for unitary concentration of probability for a sequence of measurement-driven refinements, while preserving the feedback-based and non-variational character of FALQON.

Under this geometric interpretation, the approach's performance is naturally bounded by the representational power of the chosen manifold. Specifically, it is expected to be most effective when the dominant structure of the output distribution is captured by local marginals, while strongly correlated distributions may require extensions beyond product-state initialization.

\section{Methodology and Experimental Setup}\label{sec:setup}

We validate our approach using weighted \textit{MaxCut} instances defined on random complete graphs with $8$ vertices. To eliminate the global $\mathbb{Z}_2$ symmetry inherent to the problem (bit-flip redundancy), we fix the state of a reference vertex. This constraint reduces the effective Hilbert space dimension to $n_q=7$ qubits, corresponding to $2^{7}$ possible bitstrings, while preserving the set of optimal solutions.

The diagonal cost Hamiltonian $H_p$ is constructed from Eq.~\eqref{eq:hamiltoniano_vertice_fixado}, multiplied by an overall factor $1/4$, with edge weights sampled uniformly from the interval $w_{ij}\sim\mathrm{Unif}(0,1)$. The driver Hamiltonian is chosen as the standard transverse field, $H_d=\sum_i X_i$. In the first MGI iteration, the FALQON circuit is initialized in the uniform superposition state $\ket{+}^{\otimes n_q}$, ensuring an unbiased initial distribution over all computational basis states. Throughout the study, all MGI simulations are performed using a fixed number of $2000$ measurement shots, allowing consistent comparison across configurations.

As a baseline reference, we first execute standard FALQON with $200$ layers, adopting the largest time step $\Delta t$ that preserves a monotonic decrease of the expected energy for this circuit depth. This baseline establishes the convergence behavior achievable with deep circuits and provides a reference for assessing the performance of shallow-depth MGI-FALQON implementations.

We then analyze the performance of MGI combined with FALQON using fixed values of the filtering parameter $n$. In this stage, the number of selected bitstrings $n$ is varied from $1$ to $25$, while the number of FALQON layers ranges from $2$ to $50$ in increments of $2$, using a fixed time step $\Delta t = 0.2$. For each pair of parameters, MGI is executed $100$ times, with $30$ outer iterations per run. Performance is quantified through two complementary metrics: the average probability of obtaining the optimal solution and the fraction of runs in which this probability exceeds $0.5$, providing a measure of both average performance and robustness across executions.

Next, we investigate a variable-$n$ strategy in which the number of selected bitstrings decreases along the MGI iterations. At iteration $r$, the value of $n$ is updated according to
\begin{equation}
n(r) = \left\lfloor
n_{\max} - (n_{\max} - n_{\min})
\frac{r - 1}{R - 1}
\right\rfloor,
\end{equation}
where $\lfloor \cdot \rfloor$ denotes the floor function and $R$ is the total number of MGI iterations. This expression implements a linear interpolation between $n_{\max}$ and $n_{\min}$ across the $R$ iterations. At the first iteration, $r = 1$, the fractional term vanishes and $n = n_{\max}$, while at the final iteration, $r = R$, the fraction equals one and $n = n_{\min}$. For intermediate values of $r$, $n$ decreases uniformly from $n_{\max}$ to $n_{\min}$. The floor operation ensures that the number of selected bitstrings remains integer-valued throughout the procedure.

In our simulations, we set $n_{\max}=30$ and $n_{\min}=5$. The FALQON circuit is restricted to two layers in order to remain within a shallow-circuit regime consistent with the objective of reducing circuit depth. For comparison, simulations are also performed with fixed values $n = 5$ and $n = 30$. Owing to the reduced circuit depth, a larger time step $\Delta t = 0.4$ can be employed while still ensuring a decreasing expected energy. For each configuration, we perform $100$ independent MGI executions and report the mean values and standard deviations of both the final FALQON energy and the probability of measuring the optimal solution.

To evaluate the performance of MGI across different problem instances, the algorithm is executed on a set of $1000$ independently generated graphs. For each graph, MGI is run using different FALQON depths, with $L = 5, 10, 20, 30, 40,$ and $50$ layers, and a time step $\Delta t = 0.2$, chosen to preserve the monotonic decrease of the expected energy in this regime.

Different strategies for selecting the number of retained bitstrings in the post-selection step are also evaluated. Fixed values $n = 5$, $10$, $15$, and $30$ are considered, together with adaptive strategies in which $n$ decreases linearly throughout the MGI iterations, specifically $n = 30 \rightarrow 5$ and $n = 20 \rightarrow 5$. The algorithm is further executed with different total numbers of iterations, considering $R = 10$, $30$, and $50$, allowing the influence of the iterative refinement process on performance to be systematically analyzed.

As a performance metric, we consider the mean probability of obtaining the optimal solution, averaged over the full set of $1000$ graph instances. This metric provides a direct measure of the ability of the method to concentrate probability on the configuration corresponding to the optimal solution throughout the iterative process.

In addition to the 8-vertex instances described above, we also performed simulations for larger complete graphs with 12 and 20 vertices in order to evaluate the scalability of the method. These larger instances follow the same general procedure, with adjusted time steps, filtering schedules, circuit depths, and number of measurement shots appropriate to each system size, as detailed in the corresponding results section.

\section{Results}\label{sec:results}

\begin{figure*}[htpb]
    \centering
    \includegraphics[width=1\linewidth]{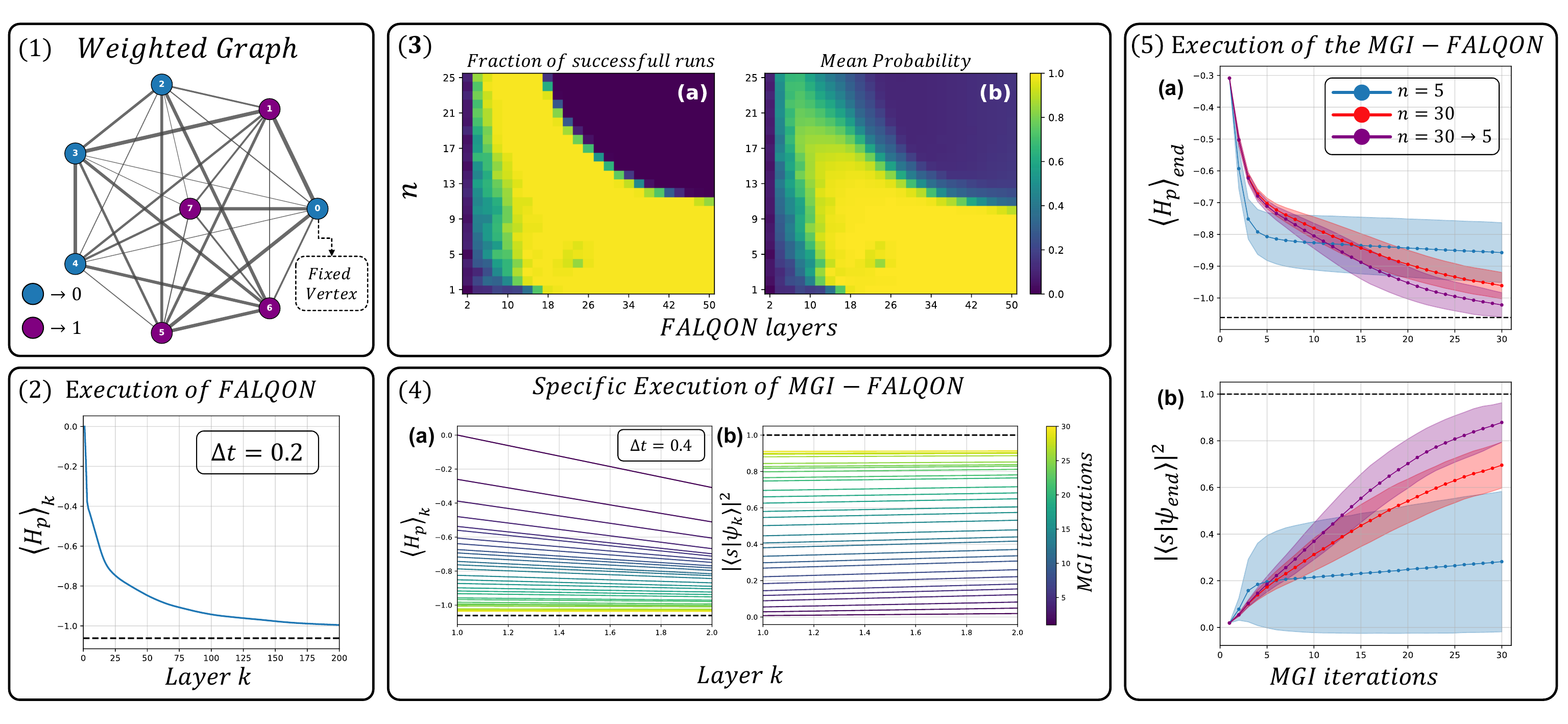}
    \caption{The figure presents, in Panel (1), the weighted graph with eight vertices considered in the MaxCut analysis, where edge weights are represented by line thickness and the vertices belonging to the two subsets of the maximum cut partition are highlighted with distinct colors, blue and purple. Panel (2) shows the evolution of the expected value of the cost Hamiltonian energy as a function of the number of FALQON layers. Panel (3) displays heat maps characterizing the performance of the MGI as a function of the number of selected bitstrings, $n$, and the number of FALQON layers. Subfigure (3a) reports the fraction of runs in which the probability of measuring the optimal solution exceeds 0.5, while subfigure (3b) presents the average probability of obtaining the optimal solution. Each point corresponds to the mean over 100 independent MGI executions. Panel (4) shows a specific execution of the MGI with a two-layer FALQON circuit, where subfigure (4a) depicts the evolution of the expected energy and subfigure (4b) the evolution of the probability of measuring the optimal solution across MGI iterations. Panel (5) summarizes the results of 100 MGI executions on the same graph, considering three strategies for $n$. Subfigure (5a) presents the final FALQON energy and subfigure (5b) the final probability of measuring the optimal solution at each MGI iteration, with curves showing the mean values together with their corresponding standard deviations.}
    \label{fig:grafico1}
\end{figure*}

The main numerical results are summarized in Fig.~\ref{fig:grafico1}, which provides both an overview of the considered instance and a comparative analysis between standard FALQON and MGI-FALQON across different regimes. The results were obtained for the weighted MaxCut instance defined on the complete graph with eight vertices. Panel~(1) shows the graph under study, where edge thickness represents the corresponding weights and the vertices belonging to the two subsets of the maximum cut partition are highlighted in blue and purple.

Panel~(2) displays the behavior of standard FALQON using a time step $\Delta t = 0.2$, chosen as the largest value that preserves a monotonic decrease of the expected energy in this regime. Although the expected value of the cost Hamiltonian decreases monotonically as the number of layers increases, approaching the optimal solution requires a large circuit depth. This behavior highlights the main limitation of standard FALQON when restricted to shallow circuits and motivates the introduction of an iterative refinement mechanism capable of compensating for limited circuit depth.

Panel~(3) presents heat maps characterizing the performance of MGI as a function of the number of selected bitstrings $n$ and the number of FALQON layers, also using $\Delta t = 0.2$. Each point corresponds to the average over 100 independent MGI executions, with 30 outer iterations per run. Subfigure~(3a) shows the fraction of runs in which the probability of measuring the optimal solution exceeds $0.5$, while subfigure~(3b) reports the average success probability. A well-defined region in parameter space is observed where MGI achieves simultaneously high success probability and a high fraction of successful runs. 

For small numbers of layers, larger values of $n$ tend to produce better results. In this regime, the state prepared by FALQON remains broadly distributed over the Hilbert space, and retaining a larger set of bitstrings in the post-selection step provides more stable statistical information for updating the initialization. As the number of layers increases, the output distribution becomes progressively concentrated around low-energy configurations, and smaller values of $n$ become more appropriate. Including too many bitstrings in this regime introduces less relevant configurations into the update of the single-qubit marginals, reducing the efficiency of the iterative refinement.

Panel~(4) illustrates a representative execution of MGI in the shallow-circuit regime, using a two-layer FALQON circuit with $\Delta t = 0.4$. The number of selected states decreases linearly from 30 to 5 throughout the iterations. A consistent decrease in the expected energy and a corresponding increase in the probability of measuring the optimal solution are observed, indicating that even with reduced circuit depth the iterative mechanism progressively guides the state toward the optimal configuration.

Panel~(5) summarizes the average behavior over 100 independent MGI executions in this shallow setting, again using two FALQON layers and $\Delta t = 0.4$, and comparing three strategies for $n$: fixed at 5, fixed at 30, and decreasing from 30 to 5. For $n = 5$, larger standard deviations and a tendency toward stagnation at higher energies and lower success probabilities are observed. For $n = 30$, the average performance improves, consistent with the heat map analysis indicating that larger values of $n$ are advantageous in shallow regimes. The variable-$n$ strategy presents the best overall behavior, showing both faster growth of the success probability and a more consistent decrease in energy across iterations, together with reduced dispersion between runs. These results indicate that a non-static choice of $n$ leads to a more stable and efficient iterative process when FALQON is restricted to few layers.

Table~\ref{tab:tabela_1000grafos} reports the mean probability of obtaining the optimal solution as a function of the number of FALQON layers $L$, the number of MGI iterations $R$, and the number of selected bitstrings $n$, averaged over a set of 1000 independently generated graphs. These results therefore characterize the typical performance of the method across a broad distribution of instances rather than a single realization.

\begin{table}[t]
\centering
\small
\setlength{\tabcolsep}{5pt}
\begin{tabular}{cccccccc}
\toprule
& &
\multicolumn{6}{c}{Mean Probability} \\
\cmidrule(lr){3-8}
$L$ & $R$ &
{$n=5$} & {$10$} & {$15$} & {$30$} & {$30\rightarrow5$} & {$20\rightarrow5$}\\
\midrule
   & 10 & 0.216 & 0.209 & 0.208 & 0.177 & 0.220 & \textbf{0.228} \\
5  & 30 & 0.267 & 0.304 & 0.335 & 0.361 & \textbf{0.365} & 0.325 \\
   & 50 & 0.302 & 0.349 & 0.390 & \textbf{0.448} & 0.433 & 0.384 \\
\hline
   & 10 & \textbf{0.403} & 0.383 & 0.356 & 0.265 & 0.367 & 0.383 \\
10 & 30 & 0.520 & 0.521 & 0.537 & 0.498 & \textbf{0.586} & 0.571 \\
   & 50 & 0.564 & 0.556 & 0.583 & 0.592 & \textbf{0.638} & 0.602 \\
\hline
   & 10 & \textbf{0.605} & 0.550 & 0.479 & 0.331 & 0.491 & 0.538 \\
20 & 30 & 0.678 & 0.697 & 0.670 & 0.502 & 0.700 & \textbf{0.713} \\
   & 50 & 0.694 & 0.722 & 0.709 & 0.552 & 0.740 & \textbf{0.743} \\
\hline
   & 10 & \textbf{0.667} & 0.592 & 0.511 & 0.397 & 0.528 & 0.579 \\
30 & 30 & \textbf{0.736} & 0.728 & 0.671 & 0.525 & 0.712 & \textbf{0.736} \\
   & 50 & 0.745 & 0.750 & 0.703 & 0.550 & 0.747 & \textbf{0.762} \\
\hline
   & 10 & \textbf{0.702} & 0.613 & 0.547 & 0.468 & 0.561 & 0.602 \\
40 & 30 & \textbf{0.768} & 0.727 & 0.666 & 0.570 & 0.713 & 0.742  \\
   & 50 & \textbf{0.776} & 0.744 & 0.688 & 0.587 & 0.749 & 0.771 \\
\hline
   & 10 & \textbf{0.727} & 0.647 & 0.596 & 0.532 & 0.604 & 0.638 \\
50 & 30 & \textbf{0.784} & 0.737 & 0.686 & 0.615 & 0.723 & 0.745  \\
   & 50 & \textbf{0.793} & 0.749 & 0.703 & 0.630 & 0.754 & 0.776  \\
\bottomrule
\end{tabular}
\caption{MGI performance as a function of the number of FALQON layers ($L$), the number of iterations ($R$), and the number of selected bitstrings ($n$). The table reports the mean probability of obtaining the optimal solution for each configuration. Bold values indicate the best performance across different bitstring selection strategies for fixed values of $L$ and $R$.}
\label{tab:tabela_1000grafos}
\end{table}

\begin{figure*}[t]
    \centering
    \includegraphics[width=0.95\linewidth]{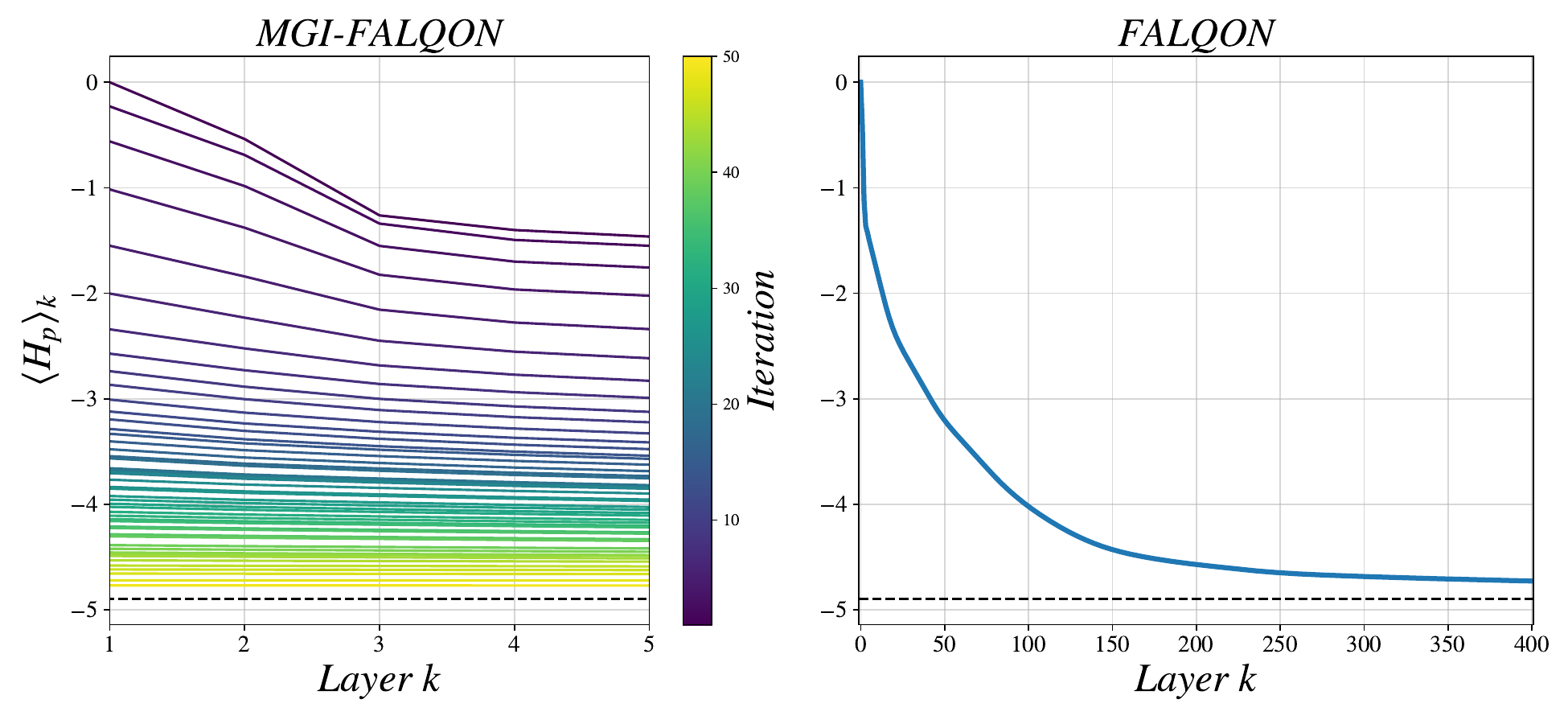}
    \caption{
Comparison between MGI-FALQON and standard FALQON for a 12-node
weighted MaxCut instance. The left panel shows the evolution of the
expected cost Hamiltonian energy $\langle H_p \rangle_k$ as a function
of the layer index for MGI-FALQON, where each curve corresponds to a
different MGI iteration. The right panel shows the corresponding
evolution for standard FALQON. The dashed horizontal line indicates the
optimal energy. While standard FALQON requires a large number of layers
to approach the optimal solution, MGI-FALQON achieves comparable energy
reduction through successive measurement-guided refinements while
keeping the circuit depth fixed.
}
    \label{fig:mgi_12nodes_layers}
\end{figure*}

\begin{figure*}[t]
    \centering
    \includegraphics[width=0.95\linewidth]{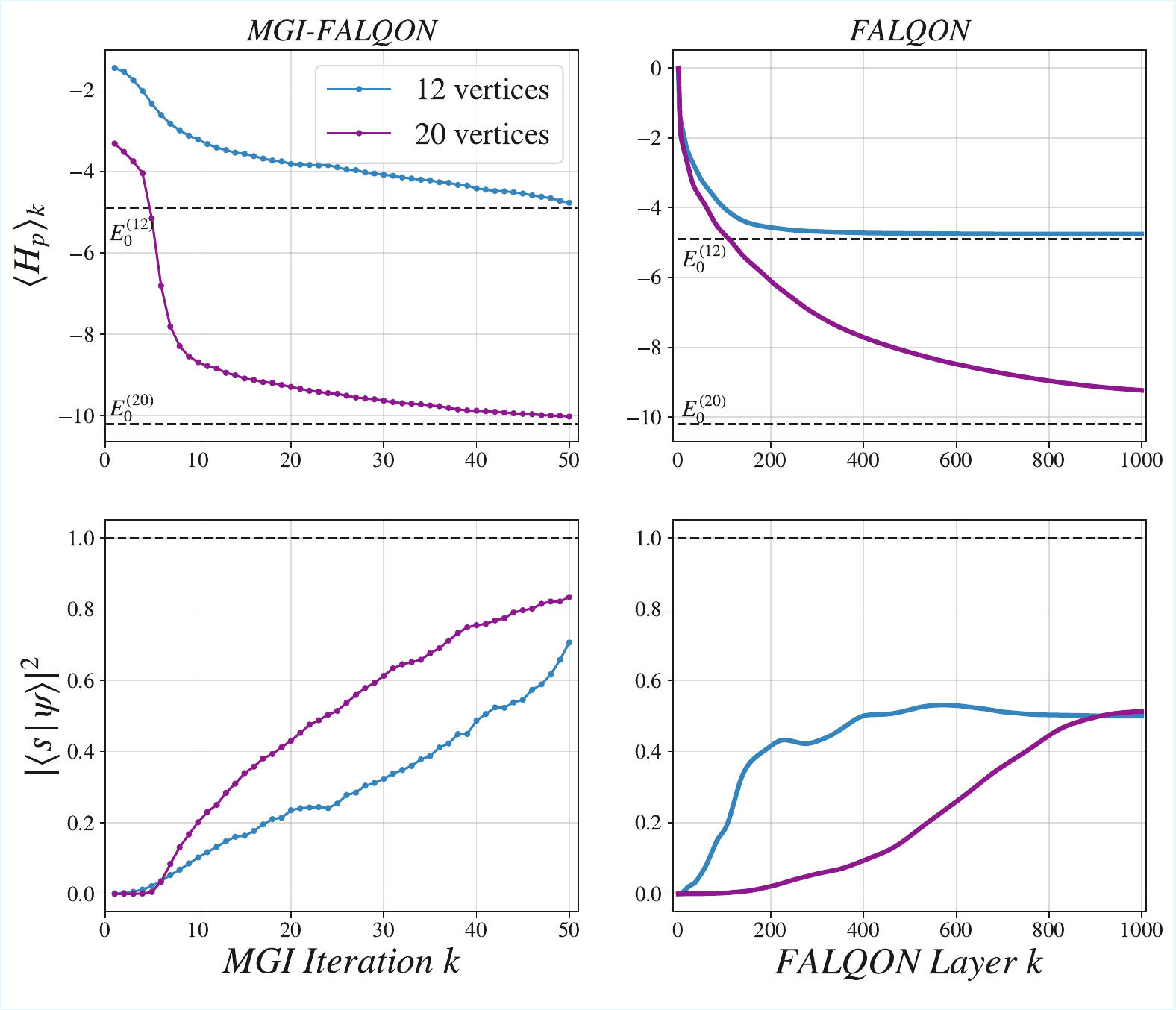}
    \caption{
Evolution of the expected energy and success probability for a
12- and 20-vertex weighted MaxCut instance. Left panels correspond to
MGI-FALQON with fixed circuit depth ($L=5$ for the 12-vertex instance em $L=20$ to the 20-vertex instance), showing the decrease of the expected energy (top) and the increase of the probability of
measuring the optimal solution (bottom) as a function of the MGI iteration. Right panels show the corresponding behavior for standard FALQON as a function of circuit depth. The dashed horizontal lines indicate the optimal values. In the energy panels, they mark the ground-state energies for each instance, $E_0^{(12)}$ for the 12-vertex case and $E_0^{(20)}$ for the 20-vertex case. In the probability panels, they indicate the value equal to one. The results illustrate that iterative
measurement-guided initialization enables shallow circuits to progressively concentrate probability on low-energy configurations, reproducing the qualitative convergence behavior of much deeper FALQON circuits.
}
    \label{fig:mgi_20nodes_iterations}
\end{figure*}

A consistent improvement in performance is observed as the number of layers $L$ increases for all strategies, reflecting the increased ability of deeper circuits to concentrate probability on configurations associated with the optimal solution. Increasing the number of iterations $R$ also leads to systematic improvements, indicating that the iterative MGI process progressively refines the initialization and reinforces the statistical evidence of favorable configurations throughout successive executions.

The choice of the number of selected bitstrings plays a central role in performance. Large values of $n$, such as $n=30$, perform competitively in very shallow circuits but lose efficiency as circuit depth increases, since the inclusion of a larger number of bitstrings introduces configurations that are less relevant once the distribution becomes concentrated. Conversely, smaller values such as $n=5$ become more effective at larger depths, where the output distribution is already localized around low-energy states.

Adaptive strategies, such as $n=30 \rightarrow 5$ and $n=20 \rightarrow 5$, consistently exhibit competitive or superior performance in shallow regimes. For example, at $L=5$ and $R=10$, the $n=20 \rightarrow 5$ strategy achieves the highest mean probability, while at $L=10$ and $R=50$ the $n=30 \rightarrow 5$ strategy yields the best result. These observations indicate that using a larger number of bitstrings during early iterations improves robustness when the circuit depth is limited, while progressively reducing $n$ increases precision as the distribution becomes more concentrated.

For deeper circuits, fixed small values such as $n=5$ tend to provide the best performance. For instance, at $L=40$ and $R=50$, the highest mean probability is obtained with $n=5$, while adaptive strategies achieve comparable but slightly lower values. A similar behavior is observed at $L=50$. Nevertheless, adaptive strategies consistently outperform fixed large values of $n$, indicating that progressively reducing $n$ maintains a balance between exploration in early iterations and refinement in later stages. Overall, these results demonstrate that the number of selected bitstrings directly influences the efficiency of the MGI process, particularly in shallow-circuit regimes where adaptive strategies provide the most stable performance.

To assess whether the qualitative behavior observed for small instances persists for larger problems, we performed additional simulations for graphs containing 12 and 20 vertices. For the 12-vertex instance, standard FALQON was executed with a fixed time step of $0.05$, while in the MGI-FALQON protocol the time step was varied across outer iterations, starting at $0.055$ and decreasing to $0.035$ at a constant rate. In addition, the number of outer iterations $n$ was progressively reduced from 200 to 5. The maximum circuit depth in the MGI-FALQON protocol was fixed at $L=5$ layers for all outer iterations. Within each iteration, FALQON was executed incrementally, with the number of layers increasing from 1 up to the fixed maximum value $L=5$. The measurement and filtering stage of MGI-FALQON was performed using 2000 shots at each iteration.

For the 20-vertex instance, standard FALQON was executed with a fixed time step $\Delta t = 0.02$. In the MGI-FALQON protocol, the time step was varied across outer iterations, starting at $\Delta t = 0.03$ and decreasing linearly to $\Delta t = 0.01$ at a constant rate. The number of outer iterations $n$ was progressively reduced from 200 to 10. The maximum circuit depth was fixed at $L=20$ layers for all outer iterations. The measurement and filtering stage was performed using 4000 shots at each iteration, ensuring sufficient statistical accuracy for the iterative update of the initial state.

Figure~\ref{fig:mgi_12nodes_layers} shows the evolution of the expected energy for the 12-node instance, comparing MGI-FALQON and standard FALQON. While standard FALQON requires a large number of layers to approach the optimal solution, measurement-guided initialization achieves a comparable reduction in energy through successive outer iterations while keeping the maximum circuit depth fixed. This behavior indicates that part of the probability concentration normally achieved through deep unitary evolution can instead be realized through iterative refinement of the initial state.

Figure~\ref{fig:mgi_20nodes_iterations} illustrates the same mechanism, comparing a 12-node instance with a larger 20-node instance, showing both the evolution of the expected energy and the probability of measuring the optimal solution. In the MGI-FALQON case, the energy shown at each iteration corresponds to the energy of the state obtained after executing FALQON with fixed maximum depth. The energy decreases monotonically across iterations, while the success probability increases steadily, reproducing the qualitative convergence pattern observed in deep FALQON circuits. Importantly, this behavior emerges even though the maximum circuit depth remains fixed, supporting the interpretation that measurement statistics can be reused to iteratively reshape the initial state toward regions of lower energy. The increase in the required fixed maximum depth (L) with system size reflects the increased expressibility required to generate informative measurement statistics, while the qualitative convergence mechanism remains unchanged.

To quantify the practical advantage of measurement-guided
initialization, we compare the resources required by standard
FALQON and MGI-FALQON to reach a fixed energy target. Since the
primary limitation of NISQ devices is the maximum achievable
coherent circuit depth, we adopt the maximum circuit depth per
execution as the relevant resource.

Let $E^\star$ denote the optimal energy for a given instance (dashed line in our figures) and
$E^{(0)}$ the energy of the initial state. We define the normalized
energy gap as
\begin{equation}
\delta(t)=\frac{\langle H_p\rangle(t)-E^\star}{E^{(0)}-E^\star},
\end{equation}
and fix a tolerance $\delta_0=0.1$, corresponding to reaching
90\% of the total achievable energy improvement.

For standard FALQON, the normalized gap decreases as the circuit
depth increases, and the relevant resource is therefore the
minimum number of layers $k_{\mathrm{target}}$ required to satisfy
$\delta \le \delta_0$. In contrast, MGI-FALQON operates with a
fixed maximum circuit depth $L$, and improvement is obtained through an
outer iteration that updates the initial state based on
measurement outcomes while keeping the circuit depth unchanged.
The comparison therefore evaluates depth efficiency rather than total computational cost, since MGI replaces circuit depth by repeated shallow executions.

Figures~\ref{fig:grafico1},  \ref{fig:mgi_12nodes_layers},  and \ref{fig:mgi_20nodes_iterations} show that comparable energy targets can be reached using shallow circuits in the MGI-FALQON protocol, whereas standard FALQON requires substantially deeper circuits to achieve the same energy level. For the representative instances considered, standard FALQON requires depths ranging
from $\mathcal{O}(10^2)$ to $\mathcal{O}(10^3)$ layers, while MGI-FALQON achieves similar energy values using fixed depths
$L=2$, $L=5$, and $L=20$ for $N=8$, $12$, and $20$ vertices, respectively.

These results indicate that part of the probability concentration normally achieved through deep unitary evolution in FALQON can be replaced by iterative measurement-guided refinement of the
initial state. In practical NISQ settings, where coherence time limits the maximum executable circuit depth, this tradeoff allows shallow circuits to approach the performance of substantially deeper evolutions without increasing circuit depth.

\section{Conclusions}
\label{sec:conclusions}

In this work, we introduced the Measurement-Guided Initialization (MGI) strategy as an iterative procedure designed to improve the performance of the FALQON algorithm in regimes where circuit depth is limited. This improvement can be interpreted as an iterative compression of measurement information into the initialization, effectively trading circuit depth for measurement-driven refinement.
The results obtained with standard FALQON confirm that, although the algorithm guarantees a monotonic decrease of the expected energy, achieving high success probabilities typically requires a large number of layers. This requirement restricts its practical applicability when circuits must remain shallow, as is currently the case in NISQ devices.

The results demonstrate that the proposed strategy improves the performance of FALQON in this regime by iteratively updating the initial state using information extracted from measurement outcomes of previous executions. This procedure progressively increases the probability associated with bitstrings that provide stronger evidence of corresponding to low-energy configurations. Even when using shallow circuits, a consistent decrease in energy and a corresponding increase in the probability of measuring the optimal solution are observed across iterations, indicating that the iterative refinement process can partially compensate for the limitations imposed by restricted circuit depth. This behavior can be interpreted as iteratively projecting measurement statistics onto a tractable product-state ansatz, thereby trading circuit depth for measurement-driven refinement.

The analysis further shows that the number of selected bitstrings plays a central role in determining performance, and that its optimal value depends on the circuit-depth regime. For shallow circuits, larger values of $n$ provide more stable marginal estimates and improve the robustness of the reinitialization process. As the number of layers increases and the output distribution becomes more concentrated around low-energy configurations, smaller values of $n$ become more effective by reducing the influence of less relevant bitstrings during the update of the single-qubit marginals.

The results also indicate that adaptive strategies, in which the number of selected bitstrings decreases throughout the iterations, exhibit stable and efficient behavior in shallow-circuit regimes. This approach allows the use of a larger number of bitstrings in early iterations, capturing broader structural information about the distribution, while later iterations benefit from a reduced value of $n$, enabling a more precise refinement of the state. As a consequence, adaptive strategies produce a more consistent evolution of both the expected energy and the success probability across iterations.

Furthermore, the analysis performed over a set of 1000 independently generated graphs shows that these behaviors remain consistent across different problem instances. Increasing the number of MGI iterations leads to progressive performance improvements, indicating that iterative refinement of the initialization constitutes an effective mechanism across a broad class of instances. Although fixed strategies with small values of $n$ achieve the best performance in deeper circuits, adaptive strategies remain competitive and consistently outperform fixed strategies with large values of $n$, particularly in shallow-circuit regimes.

These results suggest that measurement-guided initialization provides a practical mechanism for improving shallow-depth quantum optimization protocols by leveraging information contained in measurement statistics without introducing classical parameter optimization. The effectiveness of the approach is therefore expected to depend on how well the dominant structure of the output distribution can be captured by local marginals, suggesting that extensions incorporating multi-qubit correlations constitute a natural next step.
 As future work, an important next step is the implementation of MGI-FALQON on real quantum processors in order to assess its behavior under realistic noise conditions. Another natural extension is the incorporation of multi-qubit correlations in the initialization procedure beyond product states, which may further enhance performance. Finally, applying the method to larger graphs and to other optimization problems will allow a systematic investigation of its scalability and practical limits in NISQ settings.

\begin{acknowledgments}
 P.M.P. acknowledge support from FAPESP (Grant No.\ 2023/12110-7). L.A.M. Rattighieri acknowledges support from CAPES (Grant No.\ 88887.143168/2025-00). M.C.O. and F.F.F. acknowledges partial financial support from the National Institute of Science and Technology for Applied Quantum
Computing through CNPq (Process No.~408884/2024-0) and from the São Paulo Research Foundation (FAPESP), through the Center
for Research and Innovation on Smart and Quantum Materials (CRISQuaM, Process No.~2024/00998-6). F.F.F. further acknowledges support from ONR (Project No.\ N62909-24-1-2012).
\end{acknowledgments}

\bibliographystyle{apsrev4-2}
\bibliography{bibl}

\end{document}